%% file: main.tex
\documentclass[letterpaper]{article} %
\usepackage{aaai2027}  %
\usepackage[hyphens]{url}  %
\usepackage{graphicx} %
\usepackage{natbib}  %
\usepackage{caption} %
\usepackage{booktabs}
\usepackage{multirow}
\usepackage{amsmath}
\usepackage{tikz}
\usetikzlibrary{positioning,fit,arrows.meta,backgrounds,calc,plotmarks}
\title{CascadeRec: Deployed Adaptive LLM Compute Allocation for\\ Multi-Tenant Content Recommendation}

\author{
    Kai Zhao,
    Guilin Zhang\corresponding,
    Xiang Cheng,
    Xiquan Cui,
    Henry Heng,
    Xu Chu,
    Alet Blanken
}
\affiliations{
    Workday AI Research\\
    \{kai.zhao, guilin.zhang, sean.cheng, xiquan.cui, henry.heng, xu.chu, alet.blanken\}@workday.com
}

\begin{document}

\maketitle

\begin{abstract}
Large language models (LLMs) are attractive rerankers for recommendation. Running an expensive model on every request is costly, and for users with sparse interaction histories it often lowers quality. We present CascadeRec, a two-tier LLM reranking cascade deployed in production at a large financial planning and analysis (FP\&A) SaaS provider. It serves content recommendations to enterprise tenants on AWS EKS and GKE. A lightweight Tier-1 model reranks every request. A router escalates only the requests where extra inference-time compute is likely to help, and sends them to a more expensive Tier-2 procedure (self-consistency sampling, multi-path reasoning, or a larger model). The system has run in a production A/B experiment for over a month. Compared with uniform use of the expensive tier, CascadeRec has reduced inference cost by about 10\%, with minimal latency impact and no quality regression observed to date. Tenant data cannot be released, so we also evaluate the pipeline on public benchmarks (MovieLens-1M and Amazon-Electronics). Segment-level analysis with paired-bootstrap confidence intervals shows that uniform LLM reranking lowers NDCG@10 for sparse-history users by up to 30\%, and its only statistically significant gain appears on the segment with the longest histories. Escalating cold users can make the loss larger instead of repairing it. Which segments benefit from escalation depends on the domain, and this motivates the segment-gated rollout we use in production. We describe the deployment architecture, the measured benefits, and lessons learned, including inter-tenant load imbalance as a driver for cascading and the handling of LLM output-parsing failures at the serving layer.
\end{abstract}

\begin{links}
    \link{Code}{https://github.com/GuilinDev/llm-cascade-reranking}
\end{links}

\section{Introduction}
\label{sec:intro}

Enterprise software now often includes LLM-powered recommendation. A typical task is to surface the right report, driver analysis, or planning artifact to an analyst at the right time. Our production setting is a multi-tenant financial planning and analysis (FP\&A) SaaS platform. Each tenant (a customer organization) receives personalized content recommendations. Two operational facts shape the design. First, LLM inference is the largest marginal cost of the feature: reranking every request with a large model costs an order of magnitude more than with a small one. Second, tenants differ in history depth, request volume, and content complexity. A single reranker for everyone either overspends on easy requests or underserves hard ones.

This paper describes \textbf{CascadeRec}, an adaptive compute-allocation system deployed in our production recommendation path. A lightweight Tier-1 LLM reranks every request. A router then decides, per request, whether to escalate to an expensive Tier-2 procedure (self-consistency sampling, diversified multi-path reasoning, or a larger model). The system has served live tenant traffic behind an A/B experiment for over a month on AWS EKS and GKE. Models are served through Amazon Bedrock and vLLM, and the system is moving to full rollout.

The design also rests on an empirical observation: \emph{uniform LLM reranking does not help every user, and neither does escalation}. We test this on public benchmarks with controlled history sparsity. Zero-shot LLM reranking lowers ranking quality for sparse-history users, by up to $30\%$ NDCG@10 at one observed interaction on Amazon-Electronics (significant under a paired bootstrap). Its only significant gain appears on the MovieLens segment with the longest histories ($+27\%$). On Amazon with Mistral, spending more inference-time compute on the harmed users (self-consistency or multi-path reasoning) makes the loss larger. When the prompt contains little preference evidence, extra reasoning has little to work with. The segments that benefit from escalation therefore depend on the domain, and they have to be measured. Production showed a tenant-level version of the same pattern. Escalation pays off for a minority of requests, and inter-tenant load imbalance makes uniform expensive reranking hard to operate. Selective allocation thus saves cost, and it also protects the users that expensive reranking would harm.

We make four contributions, aligned with the IAAI criteria for deployed applications. First, we describe \textbf{a deployed cascade architecture} for LLM reranking in a multi-tenant SaaS product. It spans two cloud providers (EKS, GKE) and two serving stacks (Bedrock, vLLM), with routing in the request path and graceful fallback. Second, we report \textbf{measured benefits} from a month-long production A/B: about 10\% lower inference cost to date, minimal latency impact, and no observed quality regression. %
Third, we provide \textbf{a reproducible offline evaluation} on public datasets, since tenant data cannot be released. It establishes the segment-level differences that motivate selective escalation. Fourth, we report \textbf{lessons learned} on inter-tenant load imbalance, LLM output-parsing failures and fallback design, deterministic caching of Tier-1 outputs, and the path from A/B to full rollout.

\section{Application Context and Problem}
\label{sec:problem}

Our platform serves enterprise FP\&A teams. Tenants work with dashboards, reports, planning models, and analytical narratives. The recommendation surface proposes content items that match the analyst's current intent and workspace state. Request traffic is modest in aggregate ($\approx$1{,}000 requests/day), but each request is high-value. A surfaced insight can influence a planning decision, so ranking quality matters more than raw QPS.

Three properties set this setting apart from consumer-scale recommendation.
\textbf{(1) Tenant heterogeneity.} Interaction histories per user range from a handful of events (new deployments, new analysts) to rich multi-quarter trails. Cold-start is the norm for a meaningful share of traffic, not an edge case.
\textbf{(2) Load imbalance.} Request volume is highly skewed across tenants: a small fraction of tenants accounts for the majority of daily requests. %
With uniformly expensive reranking, a few heavy tenants dominate GPU spend. During their bursts, latency degrades for everyone else. This imbalance was a primary motivation for per-request compute allocation.
\textbf{(3) Cost structure.} Tier-2 inference (a larger model or multi-sample reasoning) costs roughly 3--6$\times$ Tier-1 per request in tokens and latency. At enterprise SLAs, that spend is hard to justify on every request unless it helps on every request. As we show, it does not.

\paragraph{Why an LLM reranker.}
Three features of this surface favor a language model over a ranker trained per tenant. First, the items are documents. Reports, narratives, and planning models carry titles, descriptions, and structure that an LLM can read directly, while a collaborative model sees only item IDs. Second, per-tenant interaction logs are small and catalogs grow as tenants add content. A supervised ranker per tenant would often lack training data and would need frequent retraining. Third, a zero-shot model can take new context as text without retraining. Our deployment serves zero-shot models only, for cost, governance, and model-management reasons (Section~\ref{sec:related}). The offline study in Section~\ref{sec:offline} shows the limits of this choice. When item text is short and history is thin, the LLM has less to work with than a collaborative-filtering (CF) retriever. %

The system-design question is therefore: \emph{for which requests is expensive LLM inference worth it, and how do we decide cheaply, per request, in the serving path?}

\section{The CascadeRec System}
\label{sec:method}

\begin{figure*}[t]
\centering
\includegraphics[width=0.92\textwidth]{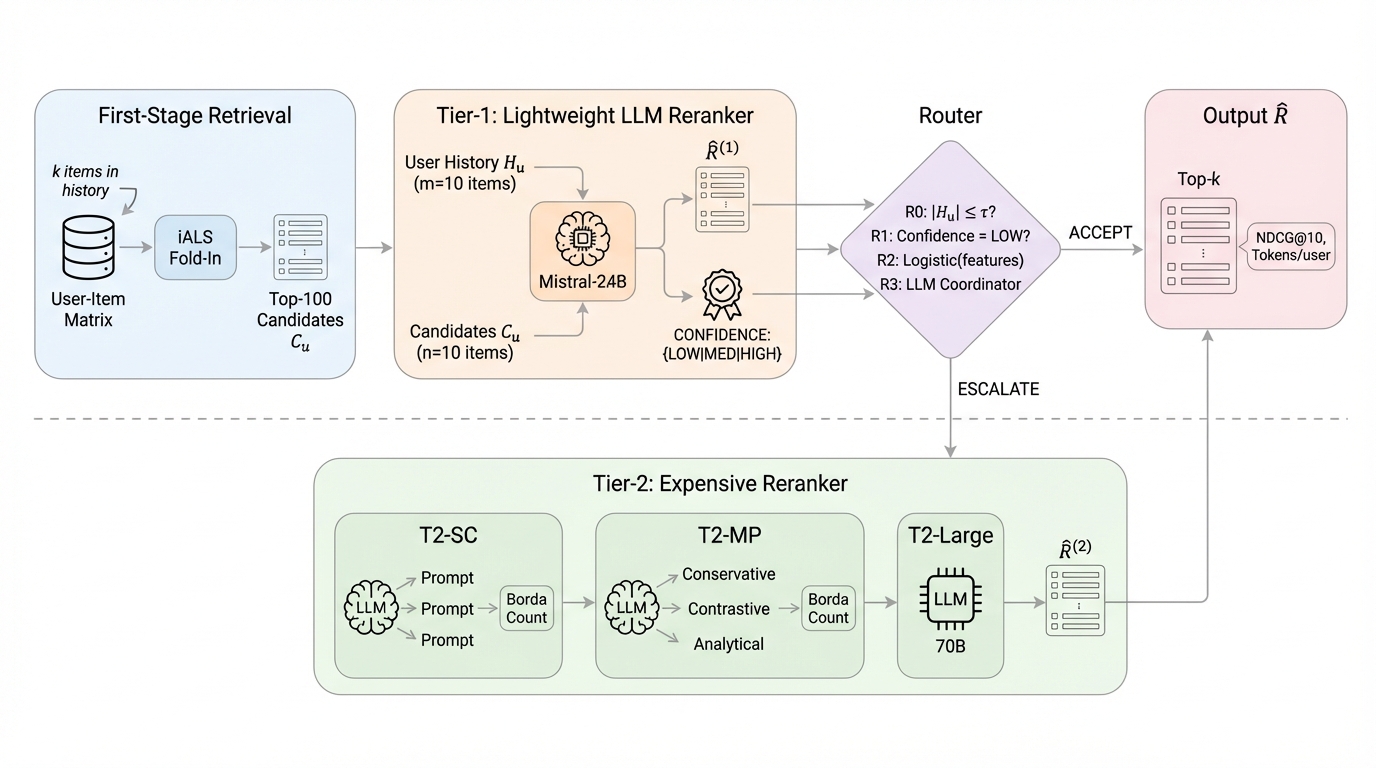}
\caption{CascadeRec architecture. Tier-1 reranks every request and emits a confidence signal. The router accepts the Tier-1 ranking or escalates the request to Tier-2.}
\label{fig:architecture}
\end{figure*}

CascadeRec is a two-tier cascade with three components (Figure~\ref{fig:architecture}). A lightweight \textbf{Tier-1} LLM reranker runs on every request. A \textbf{router} decides whether the Tier-1 result is enough. A \textbf{Tier-2} reranker applies more expensive inference only to escalated requests.

\subsection{Formulation}
Let $u$ denote a user (within a tenant), $H_u$ the interaction history, and $C_u$ the candidate set from first-stage retrieval. A cascade policy $\pi$ produces the final ranking $\hat{R}_u = \pi(H_u, C_u)$ and trades quality against inference cost:
\begin{equation}
    \max_{\pi} \sum_{u} \mathrm{NDCG@}k(\hat{R}_u) - \lambda \cdot \mathrm{cost}(\hat{R}_u).
\end{equation}

\subsection{Tier-1: Lightweight Reranking}
Tier-1 receives the user's $m$ most recent interactions and the top-$n$ retrieved candidates as text. It outputs a ranked list of candidate IDs and a coarse verbalized confidence (\texttt{LOW}/\texttt{MEDIUM}/\texttt{HIGH}). The prompt is simple on purpose: Tier-1 should be a robust, low-cost pass, and its output also serves as a routing signal. All prompts are in the released repository. Tier-1 outputs are cached per (user, model, seed). Every downstream consumer, including the A/B control arm, sees identical Tier-1 rankings. This cuts cost and makes offline and online comparisons exact.

\subsection{Routing}
We designed four router families with different cost and expressiveness. \textbf{R0} is a zero-cost heuristic on history length ($|H_u| \le \tau \Rightarrow$ escalate). \textbf{R1} uses the verbalized self-confidence of Tier-1. \textbf{R2} is a logistic classifier over cheap features. \textbf{R3} is an LLM coordinator. This paper evaluates R0 only. The offline study runs R0 with $\tau{=}5$ to test the intuitive policy of sending sparse-history users to Tier-2.

Production runs a zero-latency heuristic router, chosen for interpretability, zero added latency, and easy per-segment auditing. Its direction and threshold are configuration, not code. The offline results show that cold escalation can hurt (Section~\ref{sec:offline}), so the offline $\tau{=}5$ setting was not copied into production. The production rule is set from per-segment deltas measured on tenant traffic, and it changes only after a segment clears the rollout gate in Section~\ref{sec:lessons}.

\subsection{Tier-2: Expensive Inference}
Escalated requests receive one of three procedures. \textbf{T2-SC} is self-consistency~\cite{wang2023self}: the same prompt is sampled $N{=}3$ times and the rankings are Borda-aggregated. \textbf{T2-MP} is multi-path reasoning: three persona-diverse prompts, also Borda-aggregated. \textbf{T2-Large} uses a larger model. These match the three practical axes of inference-time scaling~\cite{snell2024scaling}: sample more, reason more diversely, or use a stronger model.

\subsection{Serving-Layer Robustness}
LLM rerankers can fail in production in ways that offline studies rarely show. We parse ranked lists with a tolerant grammar. On parse failure, the request falls back to the first-stage retrieval order, so the user never sees an error. Failures are logged per model and per segment. Hard wall-clock timeouts with mid-stream stall detection bound tail latency, and timed-out requests also fall back. In our offline matrix, parse-failure rates range from 0.1\% to 10\% depending on model and catalog (Section~\ref{sec:lessons}). Production tracks the same compliance metric per tenant segment. %

\section{Deployment}
\label{sec:deployment}

\subsection{Architecture}
CascadeRec runs on Kubernetes across AWS EKS and GKE. Model serving is split between self-hosted vLLM on GPU node pools and Amazon Bedrock managed endpoints, so each tier can be scaled and priced independently. Earlier enterprise deployments cover serverless GPU analytics~\cite{zhang2025bdaas} and predictive autoscaling of serverless workloads on Kubernetes~\cite{zhang2025aapa}. %
The router runs in the request path inside the recommendation service. Escalated requests go to the Tier-2 pool with a latency budget and fall back to the Tier-1 ranking on timeout.

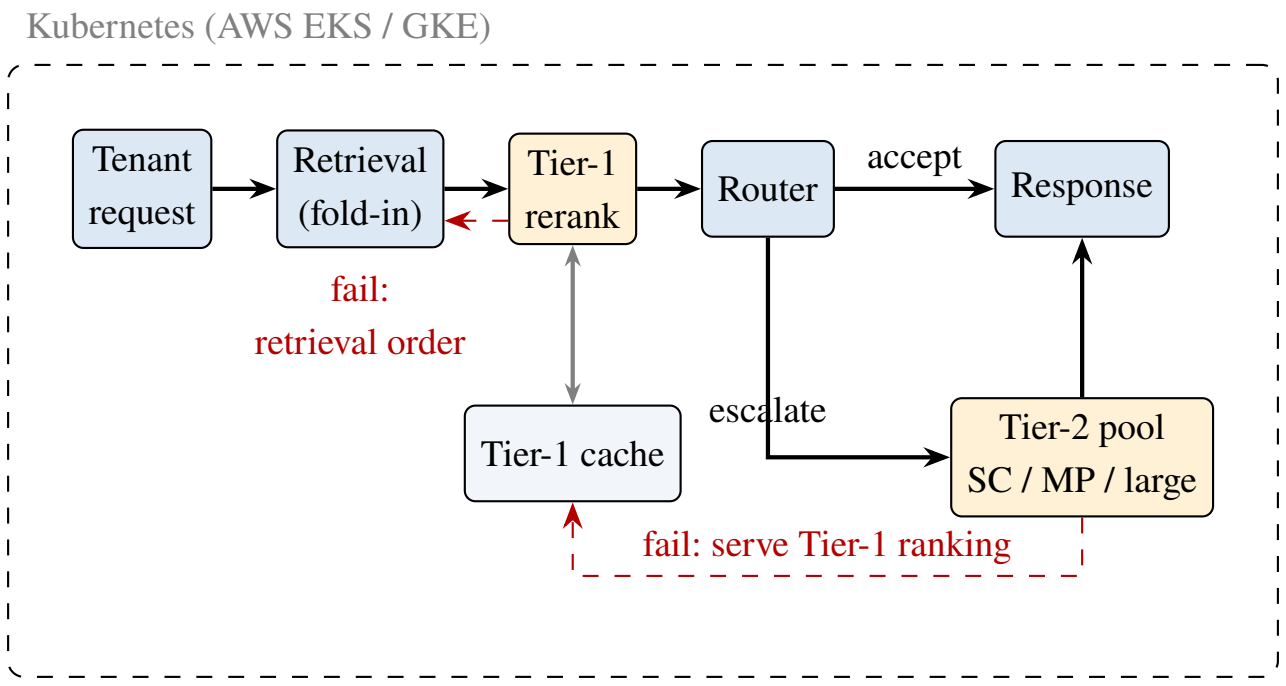
\begin{figure}[t]
\centering
\begin{tikzpicture}[
  font=\scriptsize,
  box/.style={draw, rounded corners=2pt, align=center, inner sep=3pt, minimum height=18pt},
  svc/.style={box, fill=blue!8},
  llm/.style={box, fill=orange!12},
  infra/.style={draw, dashed, rounded corners=3pt, inner sep=5pt},
  arr/.style={-{Stealth[length=5pt]}, thick},
  fb/.style={-{Stealth[length=5pt]}, dashed, red!70!black},
  node distance=14pt and 12pt
]
\definecolor{svcfill}{HTML}{DCE9F4}
\definecolor{llmfill}{HTML}{FFF1D6}
\definecolor{cachefill}{HTML}{F2F6FB}
\node[box, fill=svcfill] (req) {Tenant\\request};
\node[box, fill=svcfill, right=12pt of req] (retr) {Retrieval\\(fold-in)};
\node[box, fill=llmfill, right=12pt of retr] (t1) {Tier-1\\rerank};
\node[box, fill=svcfill, right=12pt of t1] (router) {Router};
\node[box, fill=svcfill, right=30pt of router] (out) {Response};
\node[box, fill=cachefill, below=30pt of t1] (cache) {Tier-1 cache};
\node[box, fill=llmfill, below=30pt of out] (t2) {Tier-2 pool\\SC / MP / large};
\draw[arr] (req) -- (retr);
\draw[arr] (retr) -- (t1);
\draw[arr] (t1) -- (router);
\draw[arr] (router) -- node[above=1pt, inner sep=1pt]{accept} (out);
\draw[arr] (router.south) |- node[pos=0.45, above, inner sep=2pt]{escalate} (t2.west);
\draw[arr] (t2.north) -- (out.south);
\draw[arr, gray, {Stealth[length=4pt]}-{Stealth[length=4pt]}] (t1.south) -- (cache.north);
\draw[fb] ([yshift=-6pt]t1.west) -- ([yshift=-6pt]retr.east);
\node[red!70!black, inner sep=1pt, align=center, below=4pt of retr] {fail:\\retrieval order};
\draw[fb] (t2.south) -- ++(0,-11pt) -| node[pos=0.25, above, inner sep=2pt]
      {fail: serve Tier-1 ranking} (cache.south);
\coordinate (footroom) at ($(t2.south)+(0,-18pt)$);
\begin{scope}[on background layer]
\node[infra, fit=(req)(retr)(router)(out)(t2)(cache)(footroom), inner sep=12pt,
      label={[anchor=south west, gray]north west:{Kubernetes (AWS EKS / GKE)}}] {};
\end{scope}
\end{tikzpicture}
\caption{Production request path. The router escalates a minority of requests to the Tier-2 pool. Any LLM failure falls back to the best earlier ranking (dashed red). Orange boxes are LLM inference.}
\label{fig:deployment}
\end{figure}

\paragraph{Request path.}
Figure~\ref{fig:deployment} shows the five stages of a request. (1) First-stage retrieval produces the candidate list for the user's tenant context. (2) The Tier-1 reranker receives the text history and candidates. Its output is parsed under a hard wall-clock timeout with mid-stream stall detection. (3) The router inspects the request's history length and the Tier-1 confidence signal, and either accepts the Tier-1 ranking or escalates. (4) Escalated requests fan out to the Tier-2 pool ($N{=}3$ samples for self-consistency or multi-path) under a separate latency budget, and the sampled rankings are Borda-aggregated. (5) Any failure at an LLM stage (timeout, stall, or unparseable output) degrades the request to the best earlier ranking: Tier-1 if available, otherwise retrieval order. The user-facing path never surfaces an LLM error. LLM inference runs on vLLM node pools and Amazon Bedrock endpoints. Cached Tier-1 outputs stabilize A/B comparisons and absorb retry traffic.

\paragraph{Observability.}
Each request logs, per tenant segment, the routing decision and its trigger, output compliance (parse success) per tier, token spend per tier, end-to-end and per-stage latency, and fallback events with their cause. Dashboards aggregate these logs into per-segment escalation rates, compliance rates, and cost curves. Quality deltas between arms are evaluated per history-length and tenant-size segment, not only in aggregate. The rollout gates in Section~\ref{sec:lessons} consume this segment-level view.

\subsection{A/B Experiment}
The system has run in a production A/B for over a month. The control arm receives the uniform reranking policy that preceded the cascade. The treatment arm receives CascadeRec. %

\subsection{Measurable Benefits}
\paragraph{Cost.} CascadeRec has reduced inference cost by $\approx$10\% versus the control arm to date, measured by token-level accounting. This figure compares the two arms end to end. The treatment arm contains everything that shipped with the cascade, so the figure is the saving of the deployed system and not of the routing policy alone. The offline token accounting isolates routing: in the matched offline configuration, the cascade at 60\% escalation spends 40--50\% fewer Tier-2 tokens than escalating every request. %
\paragraph{Quality.} Business-metric deltas between arms are still accumulating. No regression has been observed to date. We report parity and do not claim gains. %
\paragraph{Latency.} Impact is minimal. Non-escalated requests are unaffected, and only the routed minority of traffic pays for Tier-2 inference. %
\paragraph{Limits of the production evidence.} Three limits apply. Quality is reported as parity, without effect sizes or confidence intervals, because business metrics are still accruing. The cost figure covers the whole treatment arm, so it does not separate routing from the other changes that shipped with it. Latency impact is reported qualitatively, not as percentiles. The final A/B readout is needed to close these gaps.

\section{Offline Evaluation on Public Data}
\label{sec:offline}

Tenant data cannot be released, so we evaluate the design on public benchmarks with a reproducible pipeline. We release the code, prompts, candidate caches (including sampled user IDs), cached Tier-1 outputs, and every raw result file behind our tables (link below the abstract). All models are open-weight and served locally. The offline study asks the question that motivated the cascade: \emph{does uniform LLM reranking help all users?}

\paragraph{Setup.}
We evaluate on MovieLens-1M~\cite{harper2015movielens} and Amazon-Electronics (2023)~\cite{hou2024bridging}. Following standard cold-start practice, we truncate each user's history to $k \in \{1,3,5,10,20\}$ items (200 users per bin, 1{,}000 users per dataset). First-stage retrieval is implicit ALS~\cite{hu2008collaborative} with fold-in, so truncated users get fresh embeddings from exactly their $k$ visible interactions. The retriever sees only the truncated history, but its latent space is trained on the full population. This makes it a deliberately strong baseline. Candidate protocols differ by catalog size. On MovieLens (3.9k items), the LLM reranks the \emph{head} $n{=}10$ of the retrieved top-100. Items ranked 11--100 are outside the reranker's reach, and retriever Recall@100 is the ceiling. On Amazon (104k items after 20-core filtering of 15M reviews), full-catalog retrieval ceilings are too low for reranking evaluation. We therefore use the standard sampled protocol for LLM ranking on large catalogs~\cite{hou2024large}: the held-out target plus 19 sampled negatives, ordered by retriever score, with the LLM reranking all 20. Sampled metrics can disagree with exact metrics~\cite{krichene2020sampled}, so we read Amazon results as comparisons under this fixed protocol. We use Mistral-Small-24B and Qwen3.5-27B via Ollama with fixed seeds, and each model runs both tiers. Tier-1 outputs are cached per (user, model, seed) and reused across all methods. Non-escalated cascade requests therefore match the uniform baseline \emph{exactly}, and we verified bit-identical per-bin scores. We report NDCG@10 and token cost per user. For every delta versus B0, we report significance from a $10^4$-resample paired bootstrap over users (starred cells: 95\% CI excludes zero), plus Wilcoxon signed-rank tests.

\begin{table}[t]
\centering
\small
\caption{NDCG@10 by cold-start bin (200 users per bin). $^*$: the 95\% paired-bootstrap CI of the delta vs.\ B0 excludes zero. CR uses $\tau{=}5$, so bins with $k \le 5$ are escalated.}
\label{tab:perbin}
\setlength{\tabcolsep}{2.6pt}
\begin{tabular}{lllccccc}
\toprule
& \textbf{Model} & \textbf{Method} & $k{=}1$ & $k{=}3$ & $k{=}5$ & $k{=}10$ & $k{=}20$ \\
\midrule
\multirow{7}{*}{\rotatebox{90}{MovieLens-1M}}
& -- & B0 & .059 & .076 & .074 & .071 & .038 \\
& \multirow{3}{*}{Mistral} & B1 & .053 & .070 & .066 & .070 & .048$^*$ \\
& & CR+SC & .054 & .066 & .069 & .070 & .048$^*$ \\
& & CR+MP & .053 & .074 & .071 & .070 & .048$^*$ \\
& \multirow{3}{*}{Qwen} & B1 & .061 & .070 & .073 & .069 & .041 \\
& & CR+SC & .059 & .062$^*$ & .076 & .069 & .041 \\
& & CR+MP & .056 & .063 & .077 & .069 & .041 \\
\midrule
\multirow{7}{*}{\rotatebox{90}{Amazon-Elec.}}
& -- & B0 & .507 & .534 & .542 & .562 & .545 \\
& \multirow{3}{*}{Mistral} & B1 & .356$^*$ & .471$^*$ & .479$^*$ & .495$^*$ & .513 \\
& & CR+SC & .337$^*$ & .462$^*$ & .443$^*$ & .495$^*$ & .513 \\
& & CR+MP & .340$^*$ & .448$^*$ & .450$^*$ & .495$^*$ & .513 \\
& \multirow{3}{*}{Qwen} & B1 & .410$^*$ & .511 & .525 & .504$^*$ & .551 \\
& & CR+SC & .419$^*$ & .513 & .537 & .504$^*$ & .551 \\
& & CR+MP & .408$^*$ & .520 & .532 & .504$^*$ & .551 \\
\bottomrule
\end{tabular}
\end{table}

\begin{table}[t]
\centering
\small
\caption{Aggregate results (1{,}000 users per dataset). \textbf{Bold}: best NDCG@10 per dataset. \underline{Underline}: best within each LLM group. Esc.: share of escalated users. Tok: average tokens per user.}
\label{tab:main}
\setlength{\tabcolsep}{3.2pt}
\begin{tabular}{llcccr}
\toprule
\textbf{Model} & \textbf{Method} & \textbf{NDCG@10} & \textbf{MRR} & \textbf{Esc.} & \textbf{Tok} \\
\midrule
\multicolumn{6}{l}{\emph{MovieLens-1M}}\\
-- & B0 No-Rerank & \textbf{0.0636} & 0.0447 & 0\% & 0 \\
Mistral & B1 Uniform & 0.0612 & 0.0414 & 0\% & 1335 \\
Mistral & CR($\tau$=5)+SC & 0.0614 & 0.0415 & 60\% & 3313 \\
Mistral & CR($\tau$=5)+MP & \underline{0.0633} & 0.0440 & 60\% & 3376 \\
Qwen & B1 Uniform & \underline{0.0629} & 0.0436 & 0\% & 668 \\
Qwen & CR($\tau$=5)+SC & 0.0615 & 0.0420 & 60\% & 1693 \\
Qwen & CR($\tau$=5)+MP & 0.0613 & 0.0416 & 60\% & 1734 \\
\midrule
\multicolumn{6}{l}{\emph{Amazon-Electronics}}\\
-- & B0 No-Rerank & \textbf{0.5380} & 0.4729 & 0\% & 0 \\
Mistral & B1 Uniform & \underline{0.4628} & 0.3801 & 0\% & 2686 \\
Mistral & CR($\tau$=5)+SC & 0.4500 & 0.3798 & 60\% & 6780 \\
Mistral & CR($\tau$=5)+MP & 0.4491 & 0.3747 & 60\% & 6875 \\
Qwen & B1 Uniform & 0.5004 & 0.4214 & 0\% & 1890 \\
Qwen & CR($\tau$=5)+SC & \underline{0.5049} & 0.4272 & 60\% & 4998 \\
Qwen & CR($\tau$=5)+MP & 0.5033 & 0.4236 & 60\% & 5037 \\
\bottomrule
\end{tabular}
\end{table}

\begin{figure}[t]
\centering
\input{figures/pareto_curve.tex}
\caption{Cost and quality on MovieLens-1M: aggregate NDCG@10 against average tokens per user, with values from Table~\ref{tab:main}. The dashed line marks B0.}
\label{fig:pareto}
\end{figure}
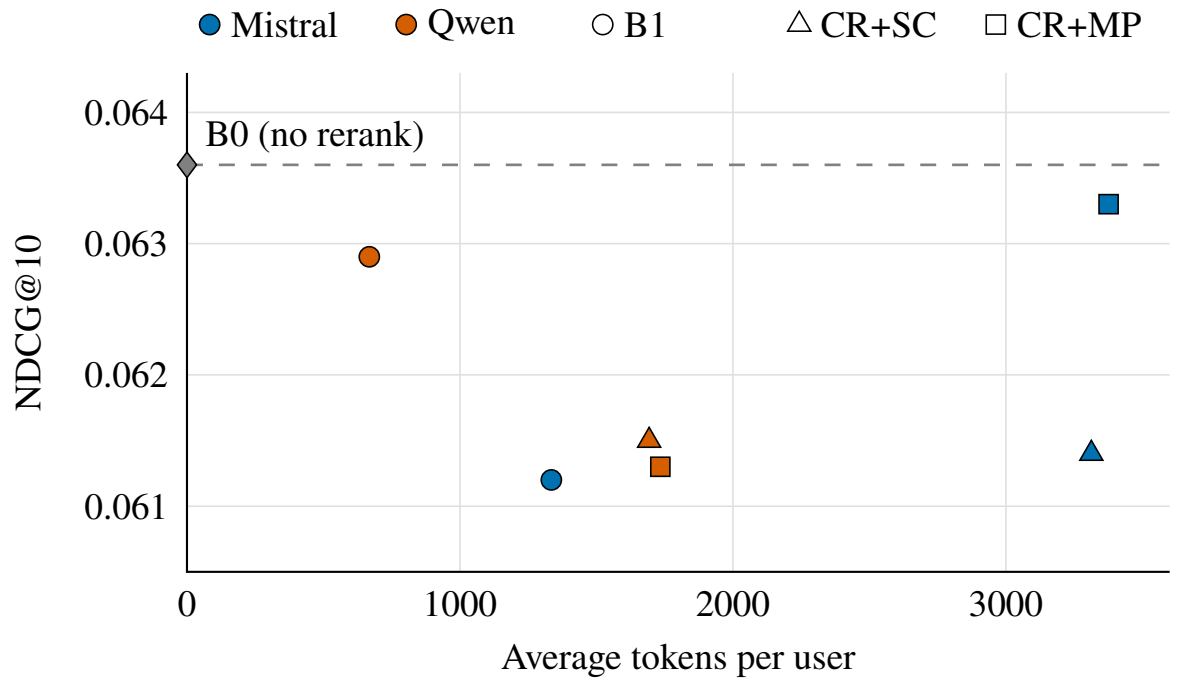

\paragraph{Uniform reranking hurts where evidence is thin.}
Table~\ref{tab:main} reports aggregate results. No LLM configuration beats the CF baseline (B0) in aggregate on either dataset. Figure~\ref{fig:pareto} plots the MovieLens configurations by cost and quality. Every LLM point lies below the B0 line, so uniform reranking (B1) and the cascade (CR) spend tokens for at best parity.
Table~\ref{tab:perbin} breaks NDCG@10 down by cold-start bin. On Amazon, uniform reranking significantly degrades sparse bins on \emph{both} models: by $29.8\%$ (Mistral) and $19.1\%$ (Qwen) relative at $k{=}1$. Mistral is significantly worse on every bin up to $k{=}10$ (aggregate $-14\%$, Wilcoxon $p<10^{-4}$). No Amazon bin improves significantly. On MovieLens the pattern is milder. Mistral degrades sparse bins by $8$--$11\%$, which is directionally consistent but individually within noise. The only statistically significant \emph{gain} anywhere appears in the bin with the longest histories ($k{=}20$: $+27\%$ Mistral, $+9.6\%$ Qwen). The harm shrinks monotonically as history grows. We attribute it to missing information. The LLM sees item titles and a short text history. The retriever encodes collaborative structure from the whole population, and the prompt does not contain that structure.

\paragraph{Escalation does not rescue sparse users, and can hurt.}
Under $\tau{=}5$ routing, users with $k \le 5$ receive $3\times$ compute through self-consistency or multi-path reasoning. On Amazon with Mistral, this lowers scores further below B1 (e.g., $k{=}1$: $.337$ vs.\ $.356$, both significantly below the $.507$ of B0). On Qwen the effect is roughly neutral. Neither Tier-2 form is better than the other. Our reading is that when the prompt lacks preference evidence, more samples or more diverse reasoning tend to reach the same wrong conclusion again. Inference-time scaling does not replace missing data. Also, the segment that gains from LLM reranking on MovieLens (long histories) is not the segment that cold escalation targets. Escalation direction must therefore be measured per segment and per domain. This is the rollout gate we use in production (Section~\ref{sec:lessons}).

\begin{table}[t]
\centering
\small
\caption{First-stage retrieval quality on MovieLens-1M (iALS fold-in). R@10 bounds what head-10 reranking can change. R@100 bounds any reranker on this candidate set.}
\label{tab:retrieval}
\begin{tabular}{rcccc}
\toprule
$k$ & \textbf{NDCG@10} & \textbf{R@10} & \textbf{R@20} & \textbf{R@100} \\
\midrule
1 & 0.059 & 0.115 & 0.205 & 0.400 \\
3 & 0.076 & 0.140 & 0.200 & 0.440 \\
5 & 0.074 & 0.140 & 0.240 & 0.480 \\
10 & 0.071 & 0.155 & 0.265 & 0.500 \\
20 & 0.038 & 0.085 & 0.155 & 0.440 \\
\bottomrule
\end{tabular}
\end{table}

\paragraph{Retrieval quality sets the ceiling.}
Table~\ref{tab:retrieval} reports the first-stage ceiling on MovieLens. On Amazon, the sampled protocol fixes the ceiling at 100\% by construction. Two observations help interpret the reranking results. First, the fold-in retriever is strong even at $k{=}1$ (NDCG@10 $0.059$ from a single observed interaction). Its user vector lives in a latent space trained on the full population, so it uses co-occurrence structure that the LLM prompt does not expose. Language-only reranking competes against information it cannot see. Second, the $k{=}20$ bin needs a cautious reading. It has the lowest CF score ($0.038$), yet it is the only bin where LLM reranking helps significantly. Users in this bin have long histories. Truncating to the 20 \emph{most recent} items before the held-out interaction over-weights recent activity, which can differ from the final click and depress the CF baseline. We therefore read the gain in this bin conservatively. LLMs add value once the text history is rich enough to compete with collaborative signal. They do not rescue users with almost no evidence, and that is the population where cold escalation spends the most and returns the least.

\paragraph{Implications for LLM reranking.}
Our production and offline evidence point to the same diagnosis: the binding constraint is missing preference evidence, not reasoning capacity. Three directions follow for practitioners. (1) \emph{Enrich the prompt, not the sampling.} Preference summaries extracted from tenant context, or item text richer than titles, change what the model knows. Repeated sampling does not. (2) \emph{Move the LLM upstream.} As a semantic candidate generator or a feature extractor for the retriever, the LLM adds signal instead of reordering a list it cannot evaluate well. (3) \emph{Fine-tune where governance allows.} Rerankers that learn collaborative patterns during training may avoid the zero-shot degradation. The cost is training pipelines and per-domain model management, which many enterprise settings (including ours) cannot justify today. A small supervised reranker is the natural baseline for this direction, and we leave that comparison to future work.

\paragraph{Scope of the offline results.}
On these public benchmarks, no cascade configuration exceeds the CF baseline in aggregate. The offline value of the cascade is to limit the harm of uniform reranking. On MovieLens its best configurations are statistically indistinguishable from B0 and keep the warm-segment gains, while spending $40$--$50\%$ fewer Tier-2 tokens than always escalating. On Amazon, all LLM configurations stay below B0. Production differs in two ways: business metrics are not NDCG, and item text is richer. There, the A/B shows the cost saving at quality parity. The offline protocol simulates history sparsity, not true inductive cold-start, because all users exist in the retriever's training matrix. We scope our claims accordingly.

\section{Deployment Insights}
\label{sec:lessons}

This section collects lessons from development, deployment, and the ongoing move from A/B to full rollout. We expect them to apply to any team that puts LLM inference into a ranking path.

\paragraph{Load imbalance is a routing problem.}
Inter-tenant load imbalance was a decisive driver. A small number of tenants generate most requests, and uniform expensive reranking turns their bursts into fleet-wide latency and cost spikes. Per-request compute allocation therefore acts as a quality policy and as an implicit load-management mechanism. Escalation budgets can be tightened under pressure without touching the always-on Tier-1 path.

\begin{table}[t]
\centering
\small
\caption{Output-compliance failure rates in the offline matrix: the share of requests whose LLM output had no parseable ranking. Such requests fall back silently.}
\label{tab:parse}
\begin{tabular}{llcc}
\toprule
\textbf{Dataset} & \textbf{Model} & \textbf{Tier-1 fail} & \textbf{Tier-2 fail} \\
\midrule
MovieLens & Mistral & 0.8--6.5\% & 2.0--3.7\% \\
MovieLens & Qwen (freeform) & 96.9\% & -- \\
MovieLens & Qwen (constrained) & 10.4\% & 5.8--7.0\% \\
Amazon & Mistral & 0.2\% & 0.2--0.3\% \\
Amazon & Qwen (constrained) & 0.1\% & 0.0--0.2\% \\
\bottomrule
\end{tabular}
\end{table}

\paragraph{Output compliance is a first-class, per-model failure mode.}
Table~\ref{tab:parse} quantifies this failure mode. Zero-shot rerankers sometimes emit unparseable output, and the rate varies by model \emph{and} domain. In our offline runs, Mistral fails to produce a parseable list on $0.2$--$6.5\%$ of requests. Under the same prompt, Qwen3.5 initially failed on \textbf{97\%} of MovieLens requests. It used its whole token budget to write recommendation essays instead of the requested ID list, so nearly every result silently fell back to retrieval order. Aggregate quality metrics did not reveal this, because the scores simply matched the no-rerank baseline. Per-request parse-rate logging did. A model-specific list-only prompt constraint reduced the failure rate to $0.1$--$10\%$. Two operational rules follow. (1) Log output-compliance rates per model and per segment as a deployment metric, next to quality. (2) Treat prompt templates as model-specific serving artifacts, not shared constants. The same trap applies to reasoning-mode APIs: Qwen's thinking mode must be disabled explicitly, or the token budget goes into hidden reasoning. With logging, silent fallback to retrieval order turns this class of outage into a quality metric we can measure.

\paragraph{Determinism is an experimentation feature.}
Caching Tier-1 outputs keyed by (user, model, seed) made A/B arms and offline replays exactly comparable. It also removed a class of spurious regressions caused by sampling noise, and it cut Tier-1 serving cost for repeated evaluations.

\paragraph{Escalation direction depends on the domain: measure it, do not assume it.}
We started this project from the intuition that cold users need more compute. On one of our two offline domains, the opposite held. On Amazon, escalating sparse-history users made their loss larger, while the only users who gained from LLM reranking anywhere were in the longest-history MovieLens bin. Had we shipped the $\tau{=}5$ cold-escalation policy to a domain like Amazon, the cascade would have spent its extra compute on the users it was hurting. A router encodes a hypothesis about \emph{where} extra inference helps. That hypothesis must be checked per segment and per domain, from measured deltas, before it controls production traffic. Routers should therefore be cheap to flip and cheap to disable, rather than fixed in the architecture.

\paragraph{Segment-level evaluation should gate rollout.}
Aggregate metrics would have approved uniform reranking, and per-segment analysis reversed that decision. Our rollout gates now require per-segment (history-length and tenant-size) deltas, not just aggregates. Full rollout proceeds cohort by cohort as each tenant cohort clears its gate. %

\section{Related Work}
\label{sec:related}

\paragraph{LLM-based recommendation.} LLMs have been used as zero-shot rankers~\cite{hou2024large,zhuang2024setwise,sun2023rankgpt}, pairwise rankers~\cite{qin2024pairwise}, two-stage rerankers~\cite{yue2023llamarec}, instruction-tuned rerankers~\cite{bao2023tallrec,luo2024recranker}, and conversational agents~\cite{gao2023chat}. LLMs also orchestrate constraint-compliant multi-agent recommendation~\cite{zhang2026dualagent}, and the robustness of LLM-generated recommendation explanations has been evaluated~\cite{zhang2026robustexplain}. Surveys cover the space~\cite{lin2025survey_llm_rec,wu2024survey_llm_rec}. Evaluation is dominated by aggregate metrics. Recent work documents cold-start biases of LLM rerankers~\cite{andre2025coldbias,lemdiasova2026diagnosing}. It also shows that LLMs underperform matrix factorization when collaborative signal reaches the model only through the prompt~\cite{pouryousef2026collabsignals}, which independently supports the cause we identify. Fine-tuned rerankers that learn collaborative patterns~\cite{bao2023tallrec,luo2024recranker} may avoid the zero-shot degradation we observe. Our deployment is limited to zero-shot serving (no per-tenant fine-tuning, for cost, governance, and model-management reasons), which is a common enterprise constraint.

\paragraph{LLM cascades and routing.} FrugalGPT~\cite{chen2023frugalgpt} introduced cascaded model selection. RouteLLM~\cite{ong2024routellm}, BEST-Route~\cite{ding2025bestroute}, Route-to-Reason~\cite{route2reason}, and budget-aware cascades~\cite{dekoninck2024cascade,zhang2024treacle} extend it. These methods route on query difficulty. Our setting routes on \emph{user information sparsity} and, operationally, on tenant load. Test-time scaling gains are known to concentrate in reasoning-heavy tasks~\cite{balachandran2025tts,zhao2025ttsknowledge}, and our results are an instance of that boundary in recommendation. Even in mathematical reasoning, measured gains depend on how the token budget is enforced and reported~\cite{zhang2026budgetboundary}.

\paragraph{Cold-start.} Classical approaches use content features, meta-learning, or contrastive methods~\cite{volkovs2017dropoutnet,zhu2020heater,wei2021clcrec,zhang2025coldstart}. LLM personalization under limited context includes LaMP~\cite{salemi2023lamp} and multi-path personalized reasoning~\cite{salemi2025pot}.

\section{Conclusion}
\label{sec:conclusion}

CascadeRec shows that adaptive, per-request LLM compute allocation is deployable in an enterprise recommendation path and pays for itself. In production it gives about 10\% lower inference cost at quality parity, with minimal latency impact and better behavior under tenant load imbalance. A reproducible offline finding motivated the design: uniform LLM reranking harms sparse-history users and helps mainly users with long histories. Production experience supports the operational corollary. Expensive inference should be spent where it helps, and the decision should be made cheaply, per request, in the serving path. We expect both the architecture and the segment-gated rollout method to transfer to other LLM-powered ranking surfaces.

\section*{Ethical Statement}
\label{sec:ethics}

Our findings have an equity implication for deployment teams. Sparse-history users are often \emph{new} users: tenants that are onboarding onto the platform, or analysts who have just joined a team. Prior work shows that recommenders already serve inactive users worse than active ones, and frames this gap as a user-side fairness problem~\cite{li2021userfairness}. Uniform LLM reranking degrades exactly this population (up to $30\%$ NDCG@10 in our offline study), while established users are unharmed or better off. Deployed naively, it would widen the quality gap between newcomers and established users, and an aggregate metric can hide that gap when established users dominate traffic. Segment-gated evaluation is therefore a fairness safeguard as well as a soundness practice, and selective routing protects the users that expensive inference would harm. The compute saving of the cascade (40--50\% fewer Tier-2 tokens than always escalating offline, $\approx$10\% of total inference cost in production) also reduces the energy footprint of LLM-based recommendation. The offline pipeline uses only public datasets with no personally identifiable information. Shared components in multi-tenant LLM systems can leak data across users~\cite{mudgal2026memleak}. Production telemetry is aggregated per segment and never leaves the tenant boundary.

\bibliography{references}

\end{document}

%% file: figures/pareto_curve.tex
% Generated by make_pareto_tikz.py from Table 2 (MovieLens-1M). Do not edit by hand.
\begin{tikzpicture}[font=\scriptsize, mk/.style={draw=black, line width=0.3pt}]
\definecolor{okblue}{HTML}{0072B2}
\definecolor{okverm}{HTML}{D55E00}
\draw[gray!25, line width=0.3pt] (0.000,0) -- (0.000,3.3);
\node[below] at (0.000,0) {0};
\draw[gray!25, line width=0.3pt] (1.806,0) -- (1.806,3.3);
\node[below] at (1.806,0) {1000};
\draw[gray!25, line width=0.3pt] (3.611,0) -- (3.611,3.3);
\node[below] at (3.611,0) {2000};
\draw[gray!25, line width=0.3pt] (5.417,0) -- (5.417,3.3);
\node[below] at (5.417,0) {3000};
\draw[gray!25, line width=0.3pt] (0,0.434) -- (6.5,0.434);
\node[left] at (0,0.434) {0.061};
\draw[gray!25, line width=0.3pt] (0,1.303) -- (6.5,1.303);
\node[left] at (0,1.303) {0.062};
\draw[gray!25, line width=0.3pt] (0,2.171) -- (6.5,2.171);
\node[left] at (0,2.171) {0.063};
\draw[gray!25, line width=0.3pt] (0,3.039) -- (6.5,3.039);
\node[left] at (0,3.039) {0.064};
\draw[line width=0.4pt] (0,3.3) -- (0,0) -- (6.5,0);
\node[below=10pt] at (3.250,0) {Average tokens per user};
\node[rotate=90] at (-1.05,1.650) {NDCG@10};
\draw[dashed, gray, line width=0.5pt] (0.000,2.692) -- (6.5,2.692);
\draw[mk, fill=gray, mark size=2.4pt] plot[mark=diamond*] coordinates {(0.000,2.692)};
\node[anchor=south west, inner sep=1pt] at (0.080,2.732) {B0 (no rerank)};
\draw[mk, fill=okblue, mark size=1.9pt] plot[mark=*] coordinates {(2.410,0.608)};
\draw[mk, fill=okblue, mark size=2.5pt] plot[mark=triangle*] coordinates {(5.982,0.782)};
\draw[mk, fill=okblue, mark size=1.8pt] plot[mark=square*] coordinates {(6.096,2.432)};
\draw[mk, fill=okverm, mark size=1.9pt] plot[mark=*] coordinates {(1.206,2.084)};
\draw[mk, fill=okverm, mark size=2.5pt] plot[mark=triangle*] coordinates {(3.057,0.868)};
\draw[mk, fill=okverm, mark size=1.8pt] plot[mark=square*] coordinates {(3.131,0.695)};
\draw[mk, fill=okblue, mark size=1.9pt] plot[mark=*] coordinates {(0.150,3.620)};
\node[anchor=west, inner sep=1pt] at (0.250,3.620) {Mistral};
\draw[mk, fill=okverm, mark size=1.9pt] plot[mark=*] coordinates {(1.450,3.620)};
\node[anchor=west, inner sep=1pt] at (1.550,3.620) {Qwen};
\draw[mk, fill=white, mark size=1.9pt] plot[mark=*] coordinates {(2.750,3.620)};
\node[anchor=west, inner sep=1pt] at (2.850,3.620) {B1};
\draw[mk, fill=white, mark size=2.5pt] plot[mark=triangle*] coordinates {(4.050,3.620)};
\node[anchor=west, inner sep=1pt] at (4.150,3.620) {CR+SC};
\draw[mk, fill=white, mark size=1.8pt] plot[mark=square*] coordinates {(5.350,3.620)};
\node[anchor=west, inner sep=1pt] at (5.450,3.620) {CR+MP};
\end{tikzpicture}

%% file: references.bib
@article{hou2024large,
  title={Large Language Models are Zero-Shot Rankers for Recommender Systems},
  author={Hou, Yupeng and Zhang, Junjie and Lin, Zihan and Lu, Hongyu and Xie, Ruobing and McAuley, Julian and Zhao, Wayne Xin},
  journal={arXiv preprint arXiv:2305.08845},
  year={2023}
}

@inproceedings{bao2023tallrec,
  title={{TALLRec}: An Effective and Efficient Tuning Framework to Align Large Language Model with Recommendation},
  author={Bao, Keqin and Zhang, Jizhi and Zhang, Yang and Wang, Wenjie and Feng, Fuli and He, Xiangnan},
  booktitle={Proceedings of the 17th ACM Conference on Recommender Systems},
  pages={1007--1014},
  year={2023}
}

@article{gao2023chat,
  title={Chat-{REC}: Towards Interactive and Explainable {LLM}s-Augmented Recommender System},
  author={Gao, Yunfan and Sheng, Tao and Xiang, Youlin and Xiong, Yun and Wang, Haofen and Zhang, Jiawei},
  journal={arXiv preprint arXiv:2303.14524},
  year={2023}
}

@article{chen2023frugalgpt,
  title={{FrugalGPT}: How to Use Large Language Models While Reducing Cost and Improving Performance},
  author={Chen, Lingjiao and Zaharia, Matei and Zou, James},
  journal={arXiv preprint arXiv:2305.05176},
  year={2023}
}

@article{ding2025bestroute,
  title={{BEST-Route}: Adaptive {LLM} Routing with Test-Time Optimal Compute},
  author={Ding, Dujian and Mallick, Ankur and Wang, Chi and Sim, Robert and Simhadri, Harsha Vardhan and Rangan, Venkatesh and Rajmohan, Saravan},
  journal={arXiv preprint arXiv:2506.22716},
  year={2025}
}

@article{route2reason,
  title={Route to Reason: Adaptive Routing for {LLM} and Reasoning Strategy Selection},
  author={Liu, Jiahao and Dong, Qiushi and Xu, Shaohan and Wang, Shuming and Ma, Yichong and Wei, Furu},
  journal={arXiv preprint arXiv:2505.19435},
  year={2025}
}

@article{dekoninck2024cascade,
  title={A Unified Approach to Routing and Cascading for {LLM}s},
  author={Dekoninck, Jasper and Baader, Maximilian and Vechev, Martin},
  journal={arXiv preprint arXiv:2410.10347},
  year={2024}
}

@article{zhang2025coldstart,
  title={Cold-Start Recommendation towards the Era of Large Language Models ({LLMs}): A Comprehensive Survey and Roadmap},
  author={Zhang, Weizhi and Bei, Yuanchen and Yang, Liangwei and Zou, Henry Peng and Zhou, Peilin and Liu, Aiwei and others},
  journal={arXiv preprint arXiv:2501.01945},
  year={2025}
}

@article{salemi2025pot,
  title={Pathways of Thoughts: Multi-Directional Thinking for Long-form Personalized Question Answering},
  author={Salemi, Alireza and Li, Cheng and Zhang, Mingyang and Mei, Qiaozhu and Li, Zhuowan and Zamani, Hamed and Bendersky, Michael},
  journal={arXiv preprint arXiv:2509.19094},
  year={2025}
}

@inproceedings{salemi2023lamp,
  title={{LaMP}: When Large Language Models Meet Personalization},
  author={Salemi, Alireza and Mysore, Sheshera and Bendersky, Michael and Zamani, Hamed},
  booktitle={Proceedings of the 62nd Annual Meeting of the Association for Computational Linguistics},
  pages={7370--7392},
  year={2024}
}

@inproceedings{hu2008collaborative,
  title={Collaborative Filtering for Implicit Feedback Datasets},
  author={Hu, Yifan and Koren, Yehuda and Volinsky, Chris},
  booktitle={2008 Eighth IEEE International Conference on Data Mining},
  pages={263--272},
  year={2008}
}

@article{harper2015movielens,
  title={The {MovieLens} Datasets: History and Context},
  author={Harper, F. Maxwell and Konstan, Joseph A.},
  journal={ACM Transactions on Interactive Intelligent Systems},
  volume={5},
  number={4},
  pages={1--19},
  year={2015}
}

@article{snell2024scaling,
  title={Scaling {LLM} Test-Time Compute Optimally can be More Effective than Scaling Model Parameters},
  author={Snell, Charlie and Lee, Jaehoon and Xu, Kelvin and Kumar, Aviral},
  journal={arXiv preprint arXiv:2408.03314},
  year={2024}
}

@inproceedings{wang2023self,
  title={Self-Consistency Improves Chain of Thought Reasoning in Language Models},
  author={Wang, Xuezhi and Wei, Jason and Schuurmans, Dale and Le, Quoc V and Chi, Ed H and Narang, Sharan and Chowdhery, Aakanksha and Zhou, Denny},
  booktitle={International Conference on Learning Representations},
  year={2023}
}

@inproceedings{zhang2026robustexplain,
  title={{RobustExplain}: Evaluating Robustness of {LLM}-Based Explanation Agents for Recommendation},
  author={Zhang, Guilin and Zhao, Kai and Friedman, Jeffrey and Chu, Xu},
  booktitle={Companion Proceedings of the ACM on Web Conference 2026},
  year={2026},
  note={arXiv:2601.19120}
}

@inproceedings{zhang2026dualagent,
  title={{LLMs} as Orchestrators: Constraint-Compliant Multi-Agent Optimization for Recommendation Systems},
  author={Zhang, Guilin and Zhao, Kai and Friedman, Jeffrey and Chu, Xu},
  booktitle={Companion Proceedings of the ACM on Web Conference 2026},
  year={2026},
  note={arXiv:2601.19121}
}

@inproceedings{zhuang2024setwise,
  title={A Setwise Approach for Effective and Highly Efficient Zero-shot Ranking with Large Language Models},
  author={Zhuang, Shengyao and Zhuang, Honglei and Koopman, Bevan and Zuccon, Guido},
  booktitle={Proceedings of the 47th International ACM SIGIR Conference},
  year={2024}
}

@inproceedings{qin2024pairwise,
  title={Large Language Models are Effective Text Rankers with Pairwise Ranking Prompting},
  author={Qin, Zhen and Jagerman, Rolf and Hui, Kai and Zhuang, Honglei and Wu, Junru and Yan, Le and Shen, Jiaming and Liu, Tianqi and Liu, Jialu and Metzler, Donald and Wang, Xuanhui and Bendersky, Michael},
  booktitle={Findings of the Association for Computational Linguistics: NAACL 2024},
  pages={1504--1518},
  year={2024}
}

@inproceedings{zhang2024treacle,
  title={Efficient Contextual {LLM} Cascades through Budget-Constrained Policy Learning},
  author={Zhang, Xuechen and Huang, Zijian and Taga, Ege Onur and Joe-Wong, Carlee and Oymak, Samet and Chen, Jiasi},
  booktitle={Advances in Neural Information Processing Systems},
  volume={37},
  year={2024}
}

@article{lin2025survey_llm_rec,
  title={How Can Recommender Systems Benefit from Large Language Models: A Survey},
  author={Lin, Jianghao and Dai, Xinyi and Xi, Yunjia and Liu, Weiwen and Chen, Bo and Zhang, Hao and Liu, Yong and Wu, Chuhan and Li, Xiangyang and Zhu, Chenxu and Guo, Huifeng and Yu, Yong and Tang, Ruiming and Zhang, Weinan},
  journal={ACM Transactions on Information Systems},
  volume={43},
  number={2},
  year={2025}
}

@article{wu2024survey_llm_rec,
  title={A Survey on Large Language Models for Recommendation},
  author={Wu, Likang and Zheng, Zhi and Qiu, Zhaopeng and Wang, Hao and Gu, Hongchao and Shen, Tingjia and Qin, Chuan and Zhu, Chen and Zhu, Hengshu and Liu, Qi and Xiong, Hui and Chen, Enhong},
  journal={World Wide Web},
  volume={27},
  number={5},
  year={2024}
}

@inproceedings{andre2025coldbias,
  title={Revealing Potential Biases in {LLM}-Based Recommender Systems in the Cold Start Setting},
  author={Andr{\'e}, Alexandre and others},
  booktitle={Proceedings of the RecSys 2025 Workshop on Evaluating and Applying Recommendation Systems with Large Language Models (EARL)},
  year={2025},
  note={arXiv:2508.20401}
}

@inproceedings{yue2023llamarec,
  title={{LlamaRec}: Two-Stage Recommendation using Large Language Models for Ranking},
  author={Yue, Zhenrui and Rabhi, Sara and de Souza P. Moreira, Gabriel and Wang, Dong and Oldridge, Even},
  booktitle={Proceedings of the CIKM 2023 Workshop on Personalized Generative AI (PGAI)},
  year={2023},
  note={arXiv:2311.02089}
}

@inproceedings{sun2023rankgpt,
  title={Is {ChatGPT} Good at Search? Investigating Large Language Models as Re-Ranking Agents},
  author={Sun, Weiwei and Yan, Lingyong and Ma, Xinyu and Wang, Shuaiqiang and Ren, Pengjie and Chen, Zhumin and Yin, Dawei and Ren, Zhaochun},
  booktitle={Proceedings of the 2023 Conference on Empirical Methods in Natural Language Processing (EMNLP)},
  pages={14918--14937},
  year={2023}
}

@article{luo2024recranker,
  title={{RecRanker}: Instruction Tuning Large Language Model as Ranker for Top-k Recommendation},
  author={Luo, Sichun and He, Bowei and Zhao, Haohan and Zhou, Yinzhi and Shao, Wei and Yao, Yuxuan and Song, Linqi and Ma, Xiaoming and others},
  journal={ACM Transactions on Information Systems},
  year={2024},
  note={arXiv:2312.16018}
}

@article{balachandran2025tts,
  title={Inference-Time Scaling for Complex Tasks: Where We Stand and What Lies Ahead},
  author={Balachandran, Vidhisha and Chen, Jingya and Chen, Lingjiao and Garg, Shivam and Joshi, Neel and Lara, Yash and Langford, John and Nushi, Besmira and Vineet, Vibhav and Wu, Yue and Yousefi, Safoora},
  journal={arXiv preprint arXiv:2504.00294},
  year={2025}
}

@article{zhao2025ttsknowledge,
  title={Test-Time Scaling in Reasoning Models Is Not Effective for Knowledge-Intensive Tasks Yet},
  author={Zhao, James Xu and Hooi, Bryan and Ng, See-Kiong},
  journal={arXiv preprint arXiv:2509.06861},
  year={2025}
}

@inproceedings{volkovs2017dropoutnet,
  title={{DropoutNet}: Addressing Cold Start in Recommender Systems},
  author={Volkovs, Maksims and Yu, Guangwei and Poutanen, Tomi},
  booktitle={Advances in Neural Information Processing Systems},
  volume={30},
  year={2017}
}

@inproceedings{zhu2020heater,
  title={Recommendation for New Users and New Items via Randomized Training and Mixture-of-Experts Transformation},
  author={Zhu, Ziwei and Sefati, Shahin and Saadatpanah, Parsa and Caverlee, James},
  booktitle={Proceedings of the 43rd International ACM SIGIR Conference on Research and Development in Information Retrieval},
  pages={1121--1130},
  year={2020}
}

@inproceedings{wei2021clcrec,
  title={Contrastive Learning for Cold-Start Recommendation},
  author={Wei, Yinwei and Wang, Xiang and Li, Qi and Nie, Liqiang and Li, Yan and Li, Xuanping and Chua, Tat-Seng},
  booktitle={Proceedings of the 29th ACM International Conference on Multimedia},
  pages={5382--5390},
  year={2021}
}

@inproceedings{ong2024routellm,
  title={{RouteLLM}: Learning to Route {LLM}s with Preference Data},
  author={Ong, Isaac and Almahairi, Amjad and Wu, Vincent and Chiang, Wei-Lin and Wu, Tianhao and Gonzalez, Joseph E. and Kadous, M. Waleed and Stoica, Ion},
  booktitle={International Conference on Learning Representations},
  year={2025},
  note={arXiv:2406.18665}
}

@article{hou2024bridging,
  title={Bridging Language and Items for Retrieval and Recommendation},
  author={Hou, Yupeng and Li, Jiacheng and He, Zhankui and Yan, An and Chen, Xiusi and McAuley, Julian},
  journal={arXiv preprint arXiv:2403.03952},
  year={2024}
}

@article{lemdiasova2026diagnosing,
  title={Diagnosing {LLM}-based Rerankers in Cold-Start Recommender Systems: Coverage, Exposure and Practical Mitigations},
  author={Lemdiasova, Ekaterina and Zmanovskii, Nikita},
  journal={arXiv preprint arXiv:2604.16318},
  year={2026}
}

@inproceedings{pouryousef2026collabsignals,
  title={Do {LLM}s Understand Collaborative Signals? Diagnosis and Repair},
  author={Pouryousef, Shahrooz and Montazer, Ali},
  booktitle={Proceedings of the 2026 ACM SIGIR Conference on Human Information Interaction and Retrieval (CHIIR)},
  year={2026}
}

@inproceedings{krichene2020sampled,
  title={On Sampled Metrics for Item Recommendation},
  author={Krichene, Walid and Rendle, Steffen},
  booktitle={Proceedings of the 26th ACM SIGKDD International Conference on Knowledge Discovery \& Data Mining},
  pages={1748--1757},
  year={2020}
}

@inproceedings{li2021userfairness,
  title={User-oriented Fairness in Recommendation},
  author={Li, Yunqi and Chen, Hanxiong and Fu, Zuohui and Ge, Yingqiang and Zhang, Yongfeng},
  booktitle={Proceedings of the Web Conference 2021},
  pages={624--632},
  year={2021}
}

@inproceedings{zhang2025bdaas,
  title={Serverless {GPU} Architecture for Enterprise {HR} Analytics: A Production-Scale {BDaaS} Implementation},
  author={Zhang, Guilin and Guo, Wulan and Tan, Ziqi and Vippagunta, Srinivas and Raman, Suchitra and Chatterjee, Shreeshankar and Lin, Ju and Liu, Shang and Schladenhauffen, Mary and Luo, Jeffrey and Jiang, Hailong},
  booktitle={2025 IEEE International Conference on Big Data (BigData)},
  pages={2503--2512},
  year={2025}
}

@inproceedings{zhang2025aapa,
  title={{AAPA}: An Archetype-Aware Predictive Autoscaler with Uncertainty Quantification for Serverless Workloads on {Kubernetes}},
  author={Zhang, Guilin and Vippagunta, Srinivas and Nandagopal, Raghavendra and Raman, Suchitra and Xu, Jeff and Pfeiffer, Marcus and Chatterjee, Shreeshankar and Tan, Ziqi and Guo, Wulan and Jiang, Hailong},
  booktitle={2025 IEEE High Performance Extreme Computing Conference (HPEC)},
  year={2025}
}

@inproceedings{zhang2026budgetboundary,
  title={Budget Boundary Effects in Test-Time Mathematical Reasoning},
  author={Zhang, Guilin and Tan, Ziqi and Guo, Wulan and Zhao, Kai and Yang, Hongyun and Luo, Mei and Ning, Qi and Yang, Feng},
  booktitle={The 6th Workshop on Mathematical Reasoning and AI (MATH-AI) at NeurIPS 2026},
  year={2026}
}

@article{mudgal2026memleak,
  title={{MemLeak}: Cross-User Semantic Leakage in Multi-Tenant {AI} Agent Memory},
  author={Mudgal, Priyanka and Zhao, Kai and Zhang, Guilin and Olsen, Andy and Miller, Ezekiel and Chu, Xu and Blanken, Aletta Johanna},
  journal={arXiv preprint arXiv:2610.04195},
  year={2026}
}
